\documentclass[12pt]{article}

\usepackage{graphicx}
\usepackage{amsmath}
\usepackage{amssymb}
\usepackage{wrapfig}
\usepackage{indentfirst}
\usepackage{color}
\usepackage{subfigure}

\begin{document}

\vskip 0.5cm \centerline{\bf\Large The reggeometric pomeron and exclusive production } 
\centerline{\bf\Large of $J/\psi$ and $\psi (2S)$  in ultraperipheral collisions at the LHC}
\vskip 0.3cm
\centerline{L\'aszl\'o~Jenkovszky $^{a\star}$, Vladyslav~Libov$^{a\diamond}$, and Magno V. T. Machado$^{b\spadesuit}$}

\vskip 1cm

\centerline{$^a$ \sl Bogolyubov ITP,
National Academy of Sciences of Ukraine, Kiev
03143 Ukraine}
\centerline{$^b$ \sl HEP Phenomenology Group, CEP 91501-970, Porto Alegre, RS, Brazil}
\vskip
0.1cm

\begin{abstract}\noindent
By using a Regge-pole model for vector meson production (VMP), successfully describing the HERA data, we analyse the correlation between VMP cross sections in photon-induced reactions at HERA and those in ultra-peripheral collisions at the Large Hadron Collider (LHC). The rapidity distributions of proton-proton collisions at 13~TeV and lead-lead collisions at 2.76 and 5.02 TeV are investigated. The transverse momentum distribution in nuclear coherent vector meson production is also  addressed.
Predictions for future experiments on production of $J/\psi$ and $\psi(2S)$ are presented.
\end{abstract}

\vskip 0.1cm

$
\begin{array}{ll}
^{\star}\mbox{{\it e-mail address:}} &
   \mbox{jenk@bitp.kiev.ua} \\
^{\diamond}\mbox{{\it e-mail address:}} &
   \mbox{vladyslav.libov@gmail.com} \\
 ^{\spadesuit}\mbox{{\it e-mail address:}} &
\mbox{magnus@if.ufrgs.br}\\  

\end{array}
$


\section{Introduction}\label{Int}

Recently, new data on $J/\psi$ production in ultra-peripheral  lead-lead (PbPb) collisions became public \cite{ALICE:2019tqa,ALICE:2021gpt,LHCb:2021bfl,ALICE:2021tyx,LHCb:2021hoq}. Interestingly, ALICE collaboration has measured for the first time the $|t|$-distribution in the coherent nuclear process $\gamma + A \rightarrow J/\psi +A $ \cite{ALICE:2021tyx}. The photo-production at low transverse momentum has been studied in peripheral PbPb collisions by LHCb collaboration at 5 TeV \cite{LHCb:2021hoq}. Vector meson production in ultra-peripheral collisions (UPC) were studied in a number of papers (see comprehensive reviews in Refs. \cite{Review,Contreras:2015dqa,Schafer2020,Klein:2020fmr}). 
Ultra-peripheral vector meson production was studied, both experimentally and theoretically in proton-proton ($pp$), proton-nucleus ($pA$) and nucleus-nucleus ($AA$) collisions at the LHC energies and below. Production of various vector mesons -- from light to heavy ones, such as $\psi (nS)$ and $\Upsilon (nS)$ states were studied. In these studies the incoming energy, rapidity and momentum transfer  of the produced system were scrutinized. These experimental studies have triggered a large number of theoretical investigations. For quarkonium production basically three classes of models exist: those based on (generalized) vector dominance model (GVDM/VDM) \cite{Klein:1999qj,Guzey:2016piu,Klusek-Gawenda:2015hja}, perturbative QCD/BFKL calculations \cite{Bautista:2016xnp,Hentschinski:2020yfm,Jones:2013pga,Jones:2013eda} and color dipole framework \cite{Goncalves:2017wgg,SampaiodosSantos:2014puz,Lappi:2010dd,Lappi:2013am,Cepila:2016uku,Cepila:2017nef,Luszczak:2019vdc,GayDucati:2018who}. The first one is quite appealing due to its simplicity and deeply connection with reggeon phenomenology and reggeon field theory \cite{Jenkovszky:2018itd}.

In this study the VMP at the CERN LHC will be scrutinized, specializing for the $J/\psi$ and $\psi (2S)$ photo-production in $pp$ and PbPb collisions. It updates our previous analysis \cite{Fiore:2014oha,Fiore:2014lxa,Fiore:2015yya} which was restricted to $pp$ collisions at 7 TeV. In Ref. \cite{Fiore:2014oha} a single-component Reggeometric Pomeron model has been considered and the focus was on the energy dependence of cross section. The inclusion of a two-pomeron (soft + hard) model and the soft dipole pomeron model was performed in Refs. \cite{Fiore:2014lxa,Fiore:2015yya}. Here, the analysis will be extended to include $pp$ collisions at 13 TeV. By using VDM approximation and Glauber multiple scattering formalism, we analyse the production in PbPb collisions at 2.76 and 5.04 TeV (predictions for XeXe collisions are also provided). The transverse momentum distribution will be investigated in the PbPb mode. 

This paper is organized as follows. In Sec. \ref{intro} we shortly review the VMP in the context of the single-component Reggeometric Pomeron model. In Subsection \ref{ppdistribution} the formalism to obtain the rapidity distribution in $pp$ collisions is shown. The numerical results both for $pp$ and PbPb reactions are compared to experimental measurements in Sec. \ref{secresults} and discussion on the theoretical uncertainties is done. In the last section we summarize the main achievements and prospects for near future are presented.

\section{Theoretical formalism}
\label{intro}
\begin{figure}[t]
\centering
\includegraphics[width=.8\textwidth]{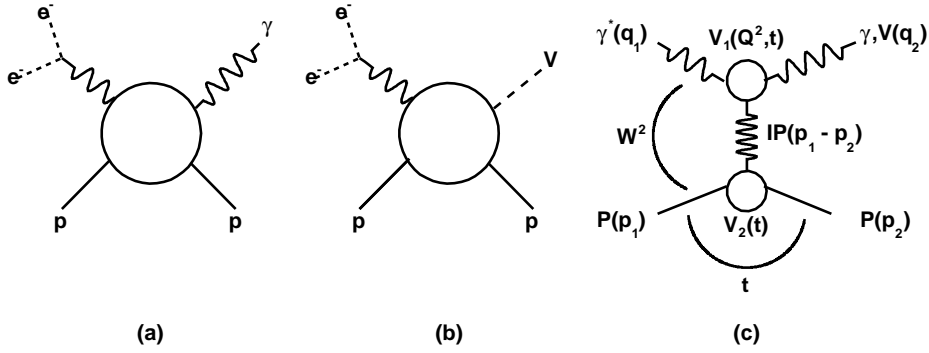}
\caption{Diagrams of DVCS (a) and VMP (b) in $e^{\pm}p$ scattering; (c) DVCS (VMP) amplitude in a Regge-factorized form (including representation for particle vertices, $V_{1,2}$).}
\label{fig:diagrams}
\end{figure}

At DESY-HERA experiment, vector meson production (VMP) was investigated in details both by the H1 and ZEUS collaborations. The same is true for the deeply virtual Compton scattering (DVCS) process. In the kinematical regime corresponding to diffractive scattering, those processes can be described by a Pomeron exchange (see Fig. \ref{fig:diagrams}). Pomeron dominance is especially clean in $J/\Psi$ production because, due to the Zweig (OZI) rule any exchange of secondary trajectories is forbidden. In this case, one has an uncontaminated Pomeron exchange which gives us the opportunity 
to study the nature of the Pomeron. An important feature is the two-fold nature of the Pomeron, which seems to be \textit{soft} or \textit{hard} depending on the virtuality, $Q^2$, of the incident photon and/or the mass of the produced vector meson, $M_V$.   
     
It is common in literature to assume that two Pomerons exist: a hard (or pQCD) Pomeron, resulting from 
perturbative Quantum Chromodynamics (QCD) calculations, and a soft one, somewhat misleadingly called \textit{non-perturbative}. 
We instead believe that there is only one Pomeron in Nature, but it has two components, whose relative weight is 
regulated by relevant $\widetilde Q^2$-dependent factors in front of the two, where the measure of the \textit{hardness}, $\widetilde Q^2=Q^2+M_V^2$, is the sum of the squared photon virtuality $Q^2$ and the squared mass $M^2_V$ of the produced vector meson.  

A specific model realizing this idea was constructed and tested against the experimental data few years ago (see
Refs.~\cite{FFJS1,FFJS2} and earlier references therein). The relevant VMP scattering amplitude reads 
\begin{eqnarray}
   A(s,t,Q^2,{M_V}^2) & = &  \widetilde{A_s}e^{-i\frac{\pi}{2}\alpha_s(t)}\left(\frac{s}{s_{0}}\right)^{\alpha_s(t)}
    e^{b_st - n_s\ln{\left(1+\frac{\widetilde{Q^2}}{\widetilde{Q_s^2}}\right)}} \nonumber \\
  & + & \widetilde{A_h}e^{-i\frac{\pi}{2}\alpha_h(t)}\left(\frac{s}{s_{0}}\right)^{\alpha_h(t)}
    e^{b_ht - (n_h+1)\ln{\left(1+\frac{\widetilde{Q^2}}{\widetilde{Q_h^2}}\right)}
    +\ln{\left(\frac{\widetilde{Q^2}}{\widetilde{Q_h^2}}\right)} },
    \label{eq:Amplitude_FFJS}
    \end{eqnarray}
 where  $\alpha_s(t)$ and $\alpha_h(t)$ are the soft and hard Pomeron trajectories, $s=W^2$, $W$ being the energy of the VMP (the center-of-mass energy of photon-nucleon system). Let us 
stress that the Pomeron is unique in all reactions, but its components (and parameters) vary. 
Examples with detailed fits can be found in the papers of Ref.~\cite{FFJS1,FFJS2}.

The integrated VMP cross section is given by 
\begin{eqnarray}
\sigma_{\gamma p\rightarrow Vp}(\widetilde Q^2, W)=\int_{t_{\mathrm{m}}}^{t_{\mathrm{thr}}}\frac{d\sigma(\widetilde Q^2, W, t)}{dt},
\label{total}
\end{eqnarray} 
where the upper limit is $t_{\mathrm{thr}} = 0$ GeV$^2$ and the lower limit is the kinematical one, $t_{\mathrm{m}} = -s/2$. The total cross section 
can be simply calculated, without integration for an exponential diffraction cone, according to the formula
\begin{equation}
\sigma_{el}(s)=\frac{1}{B(s)}\frac{d\sigma}{dt}\biggr\rvert_{t=0},
\label{elastic}
\end{equation}
where $B$ is the forward slope. 

\begin{figure}[t]
\centering
 \includegraphics[width=.4\textwidth]{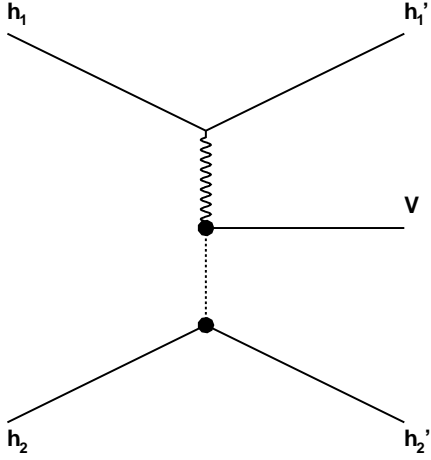}
 \caption{Feynman diagram of vector meson production in a hadronic collision.}
 \label{fig:vmp_feynman}
\end{figure}

For practical reasons, an effective Pomeron  which can be \textit{soft} or \textit{hard}, depending on kinematical domain or process defining its \textit{hardness}. In the single-component Reggeometric Pomeron the amplitude takes the simple form:
\begin{eqnarray}
\label{amplitudesingle}
A(Q^2,s,t)=\widetilde H(\widetilde{Q^2}) e^{-\frac{i\pi\alpha(t)}{2}}\left(\frac{s}{s_0}\right)^{\alpha(t)} e^{2\left(\frac{a}{\widetilde{Q^2}}+\frac{b}{2m_N^2}\right)t}, \quad \widetilde H(\widetilde{Q^2})\equiv \frac{\widetilde{A_0}}{\left(1+\frac{\widetilde{Q^2}}{{Q_0^2}}\right)^{n}},
\end{eqnarray}
where the exponent in the exponential factor in Eq. (\ref{amplitudesingle}) reflects the geometrical nature of the model. The quantities $a/\widetilde Q^2$ and $b/2m_N^2$  correspond to the effective sizes of upper and lower vertices in Fig. \ref{fig:diagrams}-c. The corresponding elastic differential cross section is given by:
\begin{eqnarray}
\frac{d\sigma_{el}}{dt}=\frac{\pi}{s^2}|A(Q^2,s,t)|^2,\quad 
\frac{d\sigma_{el}}{dt}=\frac{A_0^2}{\left(1+\frac{\widetilde{Q^2}}{{Q_0^2}}\right)^{2n}}\left(\frac{s}{s_0}\right)^{2(\alpha(t)-1)}e^{4\left(\frac{a}{\widetilde{Q^2}}+\frac{b}{2m_N^2}\right)t}.
\end{eqnarray}
Since our primary goal is the comparison between the energy (rapidity) dependence of VMP at HERA and the LHC, we start with very simple ans\"{a}tz for the $\gamma p \rightarrow  V p$ cross section, postponing the use of the more elaborated model given by Eq. (\ref{eq:Amplitude_FFJS}) to a future study. Here, we follow the studies on the single-component Reggeometric Pomeron model which has been applied to deeply virtual Compton scattering (DVCS) in Ref.~\cite{Capua}.
Apart from $W$ and $t$, it contains also the dependence on virtuality, $Q^2$.
The model was fitted to the HERA data on DVCS, but it can be applied also to the VMP by refitting its parameters.
In this paper we use the first (simpler) version of the Reggeometric model of VMP and DVCS, proposed in Refs. \cite{FFJS1,FFJS2}.

The Reggeometric Pomeron model  applies to photoproduction ($Q^2=0$) limit, where $\tilde{Q}^2=M_V^2$. The total photoproduction cross section for $J/\psi$ meson is given by  \cite{FFJS1,FFJS2}:
\begin{eqnarray}
\sigma_{\gamma p \to J/\Psi p} (W,\widetilde Q^2) & = &\frac{A_0^2}{\left(1+\widetilde Q^2/\widetilde Q^2_0\right)^{2n}}\frac{\left(\frac{W}{W_0}\right)^{4(\alpha_0-1)}}{B_V\left(W,\widetilde Q^2 \right)}, \label{eq:jpsiproduction}\\
 B_V \left(W,\widetilde Q^2 \right) & = &  4\left[\alpha'\ln(W/W_0)+\left(\frac{a}{\widetilde Q^2}+\frac{b}{2m_N^2}\right)\right].
\label{eq:bvslope}
\end{eqnarray}
The parameters, fitted \cite{FFJS1,FFJS2}  to the $J/\psi$ photoproduction data, are presented in Table \ref{tab:1}. 
Note that compared to the original formula, $s$ was replaced by $W^2$, since $W$ is used in this paper to 
denote the photon-proton centre-of-mass energy (in contrast, $\sqrt{s}$ in this paper is the proton-proton centre-of-mass energy). Our baseline model  is defined by the set of parameters fitted in Ref. \cite{FFJS1}. In next subsection we summarize the main expressions used to describe the rapidity distribution of exclusive meson production in proton-proton collisions. The input is the production  cross section presented above, Eq. (\ref{eq:jpsiproduction}).
\begin{table}[t]
  \centering
   \footnotesize
   \caption{Values of the parameters presented in Refs. \cite{FFJS1,FFJS2}  fitted to data on VMP at HERA. Our baseline prediction is based in parameter from Ref. \cite{FFJS1}.}
   \label{tab:1}
  \begin{tabular}{c||c|c|c|c|c|c|c}
     \hline
                 Ref.&$A_0$ $\left[\frac{\sqrt\text{nb}}{\text{GeV}}\right]$
                 &$\widetilde{Q^2_0}$ $\left[\text{GeV}^2\right]$&   $n$
                 &$\alpha_{0}$& $\alpha'$  $\left[\text{GeV}^{-2}\right]$
                 &$a$&$b$ \\ \hline
        \cite{FFJS1}& 29.8 $\pm$ 2.8 & 2.1 $\pm$ 0.4      &1.37 $\pm$ 0.14 &1.20 $\pm$ 0.02 & 0.17 $\pm$ 0.05& 1.01 $\pm$ 0.11 & 0.44 $\pm$ 0.08  \\ 
        \cite{FFJS2} & 30$\pm$31 &2.3$\pm$2.2        &1.45$\pm$0.32 &1.21$\pm$0.09& 0.077$\pm$0.072& 1.72       &1.16        \\  
 \hline
        \end{tabular}
 \end{table}

\subsection{Rapidity distribution in proton-proton collisions}
\label{ppdistribution}
Vector meson production cross section in hadronic collisions can be written in a factorized form in the equivalent photon approximation (EPA) \cite{Review}. The distribution in rapidity $y$ of the production of a vector meson $V$ in the reaction $h_1+h_2\rightarrow h_1+V+h_2$ (where $h_i$ stands for a hadron) is calculated according to a standard prescription based on the factorization of the photon flux and photon-proton cross section (see Fig. \ref{fig:vmp_feynman}).

Generally speaking, the $\gamma p$ differential cross section depends on three variables: the total energy   of the $\gamma p$ system, $W$, 
the squared momentum transfer at the proton vertex, $t$, and $\widetilde Q^2=Q^2+M_V^2$, where, as previously said, $Q^2$ is the photon virtuality and $M_V$ is the mass of the produced vector meson. Since, by definition, in ultra-peripheral collisions we have $b>>R_1+R_2$, where $b$ is the impact parameter, i.e. the closest distance between the centres of the colliding particles/nuclei with radii  $R_i$ (i = 1,2),
photons are nearly real, $Q^2 \simeq 0$, and $M_V^2$ remains the only measure of \textit{hardness}. Notice that this might not be true for peripheral collisions, where $b\sim R_1+R_2$, and in the 
Pomeron or Odderon exchange instead of the photon.
\begin{figure}[t]
  \centering
    \includegraphics[width=.9\textwidth]{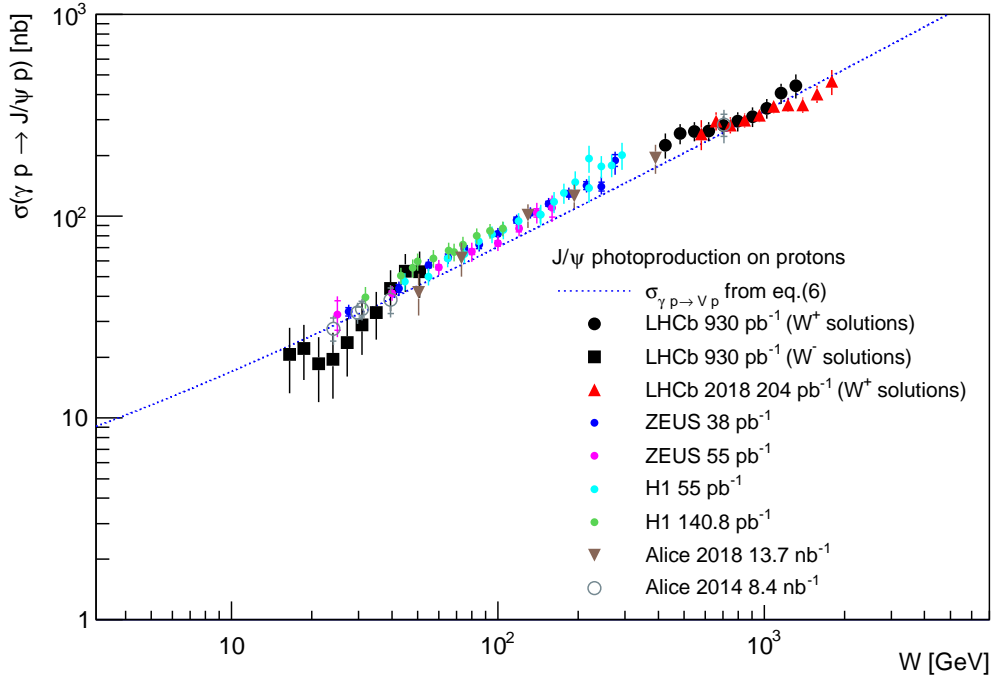}
  \caption{The cross section for exclusive $J/\Psi$ photoproduction, $\gamma p \to J/\Psi p$,  as a function of the photon-proton c.m.s. energy compared with different data sets. In addition to data considered in Refs. \cite{Fiore:2014oha,Fiore:2014lxa,Fiore:2015yya}, we added the latest ALICE \cite{ALICE:2014eof,ALICE:2018oyo} and LHCb \cite{LHCb:2018rcm} data. Dotted line is the prediction from the Reggeometric Pomeron model.}
           \label{fig:energydependence}
\end{figure}

The differential cross section, $d\sigma/dy$, as function of the meson rapidity $y$ reads
\begin{eqnarray}
\frac{d\sigma (h_1+h_2\rightarrow h_1+V+h_2)}{dy} & = & S_{\mathrm{gap}}(\omega_+)\,\omega_+\frac{dN_{\gamma/h_1}(\omega_+)}{d\omega}\sigma_{\gamma h_2\rightarrow Vh_2}(\omega_+) \nonumber \\
\label{eq:rapiditypp}
& + &S_{\mathrm{gap}}(\omega_-)\,\omega_-\frac{dN_{\gamma/h_2}(\omega_-)}{d\omega}\sigma_{\gamma h_1\rightarrow Vh_1}(\omega_-),
\end{eqnarray}
where $dN_{\gamma/h}/d\omega$ is the \textit{equivalent photon flux} \cite{Review} for protons. The quantity $S_{\mathrm{gap}}(\omega_{\pm})$ is the gap survival factor. Here, we will use the following expression for the photon flux:
\begin{equation}
\frac{dN_{\gamma/h}(\omega)}{d\omega} = \frac{\alpha_{em}}{2\pi\omega}[1+(1-\frac{2\omega}{\sqrt{s}})^2]
(\ln\Omega-\frac{11}{6}+\frac{3}{\Omega}-\frac{3}{2\Omega^2}+\frac{1}{3\Omega^3}),
\label{photon_flux}
\end{equation}
and $\sigma_{\gamma p\rightarrow Vp}(\omega)$ is the total cross section of the vector meson photoproduction sub-process. 
Here $\omega$ is the photon energy, $\omega=W^2_{\gamma p}/2\sqrt s$, where $\sqrt s$ denotes now the c.m.s. energy of the  proton-proton system; 
$\omega_{min}=M_V^2/(4\gamma_Lm_p),$ where $\gamma_L=\sqrt s/(2m_p)$
is the Lorentz factor (Lorentz boost of a single beam). Furthermore,
$\Omega=1+Q_0^2/Q_{min}^2,$ with  $Q_{min}^2=(\omega/\gamma_L)^2$ and  $Q_0^2=0.71 GeV^2$. The photon energy can be written in terms of the meson rapidity as $\omega = \frac{M_v}{2}e^y$.

In Refs. \cite{Fiore:2014oha,Fiore:2014lxa,Fiore:2015yya}, we have addressed the rapidity distribution for $J/\psi$ and $\psi (2S)$  in $pp$ collisions at the LHC for the energy of 7 TeV. Comparison of different models for the scattering amplitude has been done, where an overall gap survival factor $S_{\mathrm{gap}}=0.8$ was utilized. In particular, analysis was done using the Reggeometric Pomeron model, a soft+hard Pomeron  and the soft dipole Pomeron model as well. Moreover, the energy dependence of the extracted cross section using hadronic $pp$ scattering was scrutinized. For simplicity, we restrict ourselves to the Reggeometric Pomeron model in present analysis. In order to update the studies done in  Refs. \cite{Fiore:2014oha,Fiore:2014lxa,Fiore:2015yya} in Fig. \ref{fig:energydependence} is presented the cross section $\sigma_{\gamma p \to J/\Psi p} (W,\widetilde Q^2)$ as a function of $W$. From experimental side we added recent data from ALICE \cite{ALICE:2014eof,ALICE:2018oyo} and LHCb \cite{LHCb:2018rcm}. The result is compared to our prediction in  Fig.~\ref{fig:energydependence} (dotted line). Also the ZEUS and H1 data are shown. 

In what follows we investigate the rapidity distribution for $J/\psi$ and $\psi (2S)$ in proton-proton collisions at 13 TeV. Moreover, we address the rapidity and transverse momentum distributions in PbPb ultraperipheral collisions at 2.76 and 5.02 TeV.  Nuclear effects are described by vector dominance model and Glauber approach for multiple scattering.

\section{Results and discussions}
\label{secresults}

\begin{figure}[t]
\centering
 \includegraphics[width=.6\textwidth]{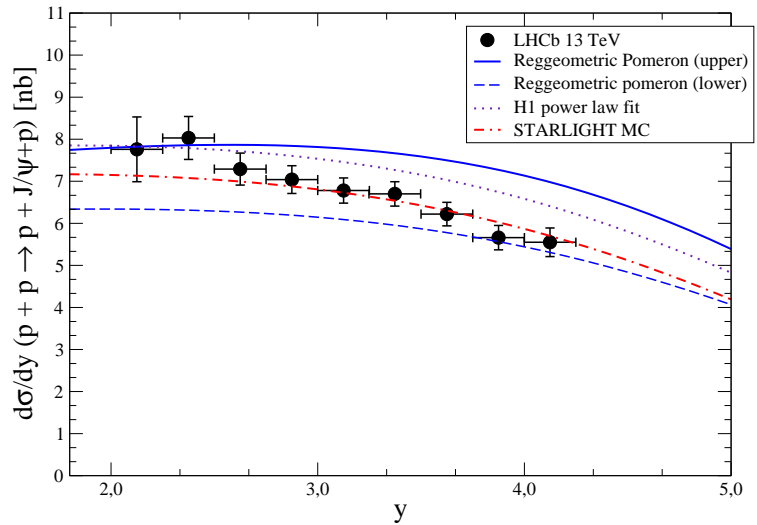}
 \caption{Rapidity distribution of exclusive $J/\psi$ production in $pp$ collisions at the LHC  ($\sqrt{s_{pp}} = 13$ TeV). Predictions of Reggeometric Pomeron are represented by solid and dashed curves, respectively.  Comparison is done with the H1 power-like fit (dotted line) and STARLIGHT fit (dot-dashed). Data from LHCb collaboration \cite{LHCb:2018rcm}.}
 \label{fig:comps13tev}
\end{figure}
\begin{figure}[t]
\centering
 \includegraphics[width=.6\textwidth]{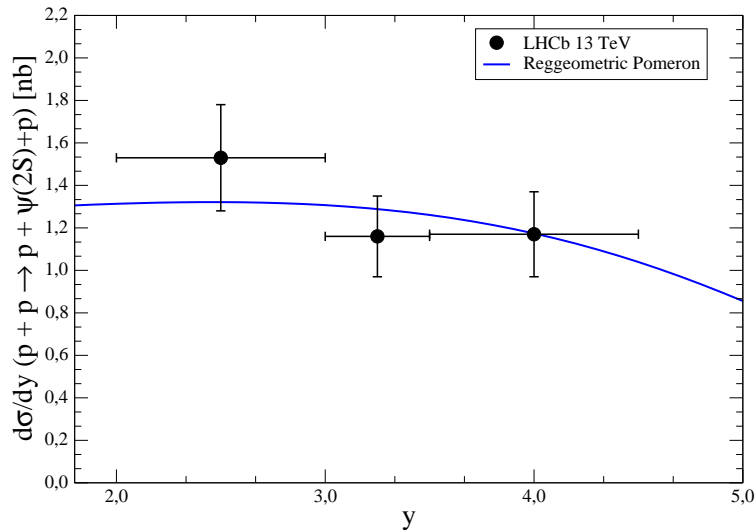}
 \caption{Rapidity distribution of $\psi (2S)$ production in $pp$ collisions at the LHC ($\sqrt{s_{pp}} = 13$~TeV). The curve stands for the  Reggeometric Pomeron model. Data from LHCb collaboration \cite{LHCb:2018rcm}.}
  \label{fig:psi2s13tev}
\end{figure}

In this section the numerical results are presented and compared to recent experimental measurements. Let us start with the rapidity distribution of $J/\psi$ and $\psi (2S)$ production in proton-proton collisions  at the LHC (see Eq. (\ref{eq:rapiditypp})). In Fig. \ref{fig:comps13tev} the predictions are shown for $J/\psi$ production at 13 TeV against the LHCb data \cite{LHCb:2018rcm}. The results for the Reggeometric Pomeron model are represented by the solid and dashed curves, respectively. The upper curve (solid) is the prediction using the values for the parameters determined in Ref. \cite{FFJS2}. On the other hand, the lower curve (dashed) corresponds to input presented in Ref. \cite{FFJS1} (see Table \ref{tab:1}). The $y$-dependent gap survival probabilities, $S_{\mathrm{gap}}(\omega_{\pm})$, are taken from Ref. \cite{Jones:2016icr}. In order to compare our prediction to other approaches in literature, it is shown also the result using the H1 power-law fit \cite{H1:2013okq} to exclusive photoproduction, $\sigma (\gamma p\rightarrow J/\psi) = \sigma_0 (W/W_0)^{\delta}$ (with $\sigma_0 = 81$ nb, $W_0= 90$ GeV and $\delta = 0.67$). This is labeled by the dotted curve in figure. Moreover, a similar fit including threshold corrections to the production cross section provided in the STARLIGHT Monte Carlo \cite{Klein:2016yzr} is also presented (dot-dashed curve). Now, the fit takes the form $\sigma (\gamma p\rightarrow J/\psi) = \sigma_0 (W/W_0)^{\delta}(1-x_V)^2$ (with $\sigma_0 = 4.06$ nb, $W_0= 1$ GeV and $\delta = 0.65$). Here, $x_V = (m_p+m_{J/\psi})^2/W^2$. The predictions using the Reggeometric Pomeron are consistent with experimental results. Of course, there is room for future adjustment of the parameters using the high precision HERA data and threshold correction, which are important at very large $y$ in the $\omega_-$ contribution.  In Fig. \ref{fig:psi2s13tev}, the rapidity distribution for $\psi (2S)$ production is presented. We have multiplied the $J/\psi$ production cross section by the measured value for the ratio $R_{\psi}=\sigma (\psi(2S))/\sigma (J/\psi)$ at very low $Q^2$. Namely, we have used $R_{\psi} = 0.17$   \cite{ZEUS:2016awm} and the upper limit for the model. The corresponding data description is reasonable given the simplification done.

Let us move now to the exclusive $J/\psi$ and $\psi (2S)$ production in PbPb collisions. In the calculations we employ traditional vector dominance model (VDM) and the classical mechanics Glauber formula for multiple scattering of vector meson in the
nuclear medium \cite{Bauer:1977iq}. In this approach, the differential cross section for the process, $\gamma +A \rightarrow J/\psi+A$ (nuclear coherent scattering), is given by:
\begin{eqnarray}
\left. \frac{\mathrm{d}\sigma\left( \gamma + A \to J/\psi+ A \right)}{\mathrm{d}t}\right|_{t=0}
& = &  \frac{\alpha_{em} \sigma_{tot}^2(J/\psi A)}{4 f^2_{J/\psi}}, \\
\sigma_{tot} \left(J/\psi A\right) & = &  \int \mathrm{d}^2 \textbf{b} 
\left[ 1-\exp\left( -\sigma_{tot}\left( J/\psi p \right) T_A\left(\textbf{b} \right) \right) \right],
\label{glauberVMD}
\end{eqnarray}
where $T_A(b)$ is the nuclear thickness function and $f_{J/\Psi}$ is the vector-meson coupling (where $f^2_{J/\Psi}/4\pi = 10.4$). The approach above has been extended in the generalized vector dominance model (GVDM) (see for instance the comprehensive review of Ref.  \cite{Frankfurt:2003wv}). The input for the Glauber model calculation in Eq. (\ref{glauberVMD}) is the cross section for the process $J/\psi+p\rightarrow J/\psi+p $ is given by:
\begin{eqnarray}
\sigma_{tot}\left( J/\psi p \right)= \left[\frac{4f^2_{J/\psi}}{ \alpha_{em}} \left.
\frac{\mathrm{d}\sigma\left( \gamma + p \to J/\psi + p
  \right)}{\mathrm{d}t}\right|_{t=0}\right]^{1/2}.
\end{eqnarray}

\begin{figure}[t]
\centering
 \includegraphics[width=.6\textwidth]{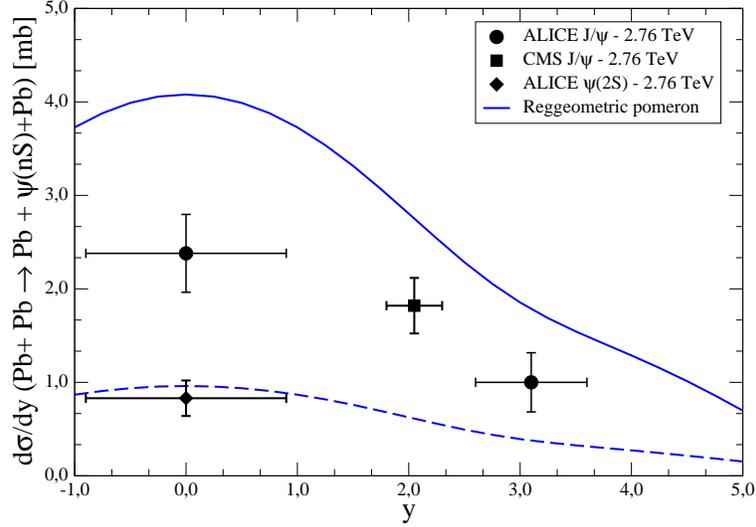}
 \caption{Rapidity distribution of $J/\psi$ and $\psi (2S)$ production in PbPb collisions at the LHC ($\sqrt{s_{\mathrm{NN}}} = 2.76$ TeV). Curves represent prediction of  the  Reggeometric Pomeron model. Data from ALICE and LHCb collaborations \cite{ALICE:2013wjo,ALICE:2012yye,ALICE:2015nmy,CMS:2016itn}.}
  \label{fig:VMAA276TeV}
\end{figure}

The total cross section for $\gamma +A \rightarrow J/\psi + A$ process can be written as:
\begin{eqnarray}
\sigma_{\gamma + A \rightarrow J/\psi + A} = 
\left. \frac{d\sigma (\gamma + A \rightarrow  J/\psi + A)}{dt}\right|_{t=0}
\int\limits_{t_{min}}^{\infty} \mathrm{d}|t| \, \left| F_A\left(t\right) \right|^2, 
\label{eq:sigtotgammaA}
\end{eqnarray}
where $F_A$ is the nuclear form factor which is the Fourier transform of the nuclear density profile,  $F_A(k^2) = \int d^3\vec{r} \, e^{ i\vec{k} \cdot \vec{r} } \rho_{A}( \vec{r} )$. In our calculation we consider an analytic form factor given by a hard sphere of radius, $R_A =1.2A^{1/3}$ fm, convoluted with a Yukawa potential with range $a$ ($a = 0.7$ fm) \cite{Davies:1976zzb}:
\begin{eqnarray}
 F_A(|k|)  =  \frac{4\pi\rho_0}{A |k^3|} \left( \frac{1}{1+a^2k^2} \right) \left[ \sin{(|k| R_A)} - |k| R_A\cos{(|k| R_A)}  \right], 
\end{eqnarray}
where $A$ is the mass number, $k$ is the momentum transfer ($k^2= |t|)$ and $\rho_0 = A/\frac{4}{3}\pi R_A^3$. 

The coherent production of $J/\psi$ in $AA$ collisions is easily computed in the equivalent photon approximation (EPA). The rapidity distribution is given by:
\begin{eqnarray}
\frac{d\sigma (A+A\rightarrow A+V+A)}{dy}=\omega_+\frac{dN_{\gamma/A}(\omega_+)}{d\omega}\sigma_{\gamma A\rightarrow VA}(\omega_+) +\omega_-\frac{dN_{\gamma/A}(\omega_-)}{d\omega}\sigma_{\gamma A\rightarrow VA}(\omega_-),
\end{eqnarray}
where the rapidity $y$  of the vector meson $V$ is related to the center-of-mass energy of the photon-nucleus system, $W_{\gamma A}^2 =\sqrt{s_{\mathrm{NN}}}M_Ve^{Y}$. The quantity $\sqrt{s_{\mathrm{NN}}}$ is the center-of-mass energy per nucleon pair in the AA system and $\omega_{\pm} = \frac{M_V}{2}e^{\pm Y}$. For simplicity, we use the analytical  photon flux corresponding to a point-like charge \cite{Klein:1999qj,Klein:2020fmr}:
\begin{eqnarray}
\frac{\mathrm{d}N_{\gamma/A}(\omega)}{\mathrm{d}\omega} = 
\frac{2Z^2 \alpha_{em}}{\pi \omega}
\left( X K_0(X) K_1(X) - \frac{X^2}{2} \left[ K_1^2(X) - K_0^2(X) \right] \right) ,
\end{eqnarray} 
where $Z$ is the nucleus atomic number, $K_{0,1}$ are the modified Bessel function and  $X = 2R_A\omega/\gamma_L$.
\begin{figure}[t]
\centering
 \includegraphics[width=.6\textwidth]{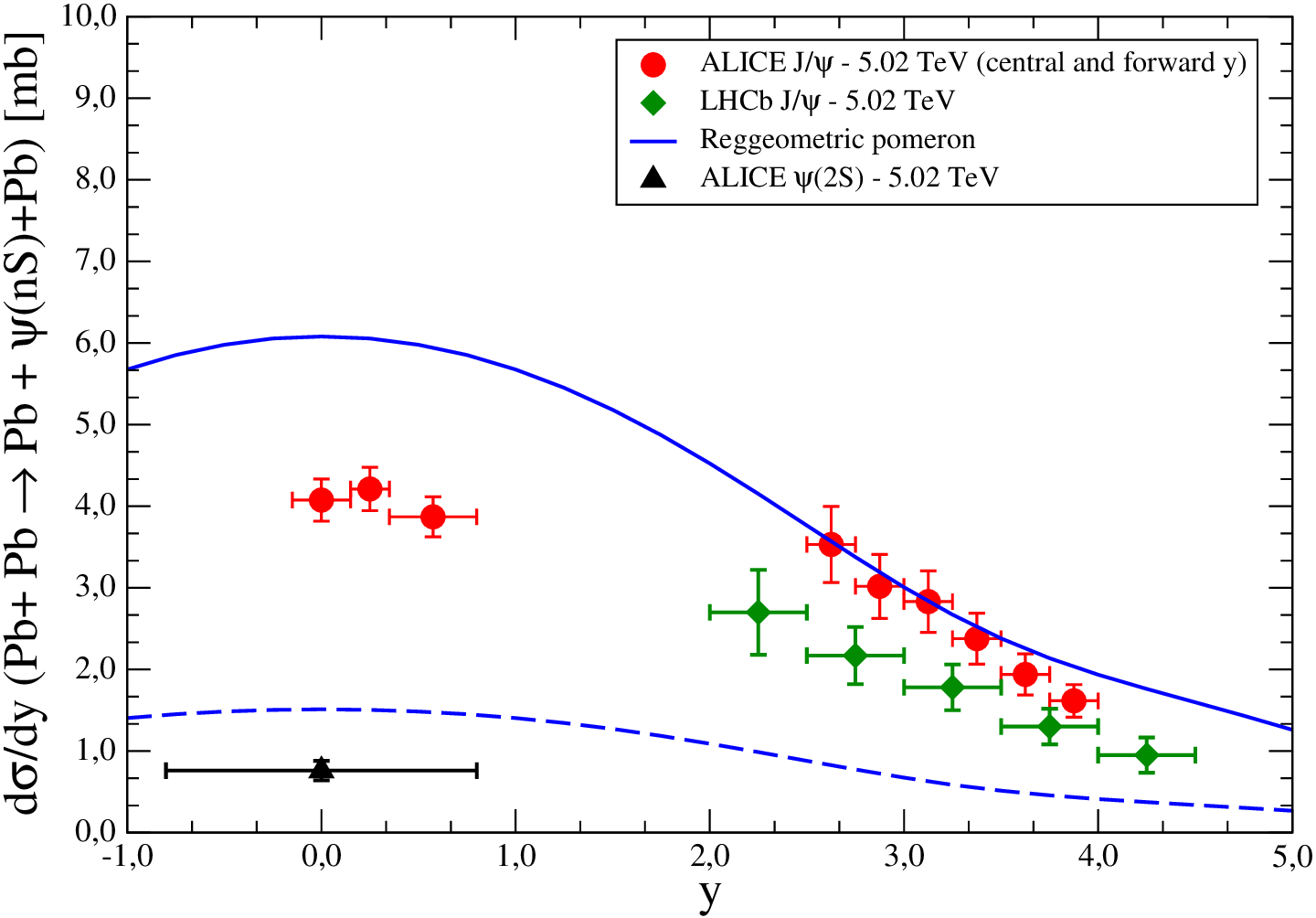}
 \caption{Rapidity distribution of $J/\psi$ and $\psi (2S)$ production in PbPb collisions at the LHC ($\sqrt{s_{\mathrm{NN}}} = 5.02$ TeV). Curves represent predictions of  the  Reggeometric Pomeron model. Data from ALICE and LHCb collaborations \cite{ALICE:2019tqa,ALICE:2021gpt,LHCb:2021bfl}.}
  \label{fig:VMAA502TeV}
\end{figure}

In Fig. \ref{fig:VMAA276TeV} results are shown for exclusive $J/\psi$ and $\psi (2S)$ production for PbPb collisions at $\sqrt{s_{\mathrm{NN}}} = 2.76$ TeV. The experimental measurements from ALICE \cite{ALICE:2013wjo,ALICE:2012yye,ALICE:2015nmy} and CMS \cite{CMS:2016itn} collaborations are presented. The prediction for the Reggeometric Pomeron are labeled by the solid ($J/\psi$) and dashed ($\psi (2S)$) curves, respectively. The input for the calculation is Eq. (\ref{eq:jpsiproduction}) using parameters of Ref. \cite{FFJS1} (see Table \ref{tab:1}). For $J/\psi$ the result overestimates the experimental measurements. This is expected as the value of $\sigma (J/\psi p)$ is small which leads to the rather small nuclear suppression. This feature is discussed in details in Refs. \cite{Guzey:2013xba,Guzey:2020ntc}. In the case of $\psi (2S)$ production the agreement with the data is reasonable.

 In Fig. \ref{fig:VMAA502TeV} results are shown for the energy of  $\sqrt{s_{\mathrm{NN}}} = 5.02$ TeV. The experimental measurements from ALICE \cite{ALICE:2019tqa,ALICE:2021gpt} and LHCb \cite{LHCb:2021bfl} collaborations are presented. 
 As observed, there is some  indication that the ALICE
and LHCb data at forward rapidity are mutually inconsistent. The Reggeometric Pomeron model is in agreement with ALICE data at forward rapidities and overestimates data at mid-rapidity. The same occurs for $\psi (2S)$ at central rapidity. The predictions follow the trend observed at 2.76 TeV.

\begin{figure}[t]
\centering
 \includegraphics[width=.6\textwidth]{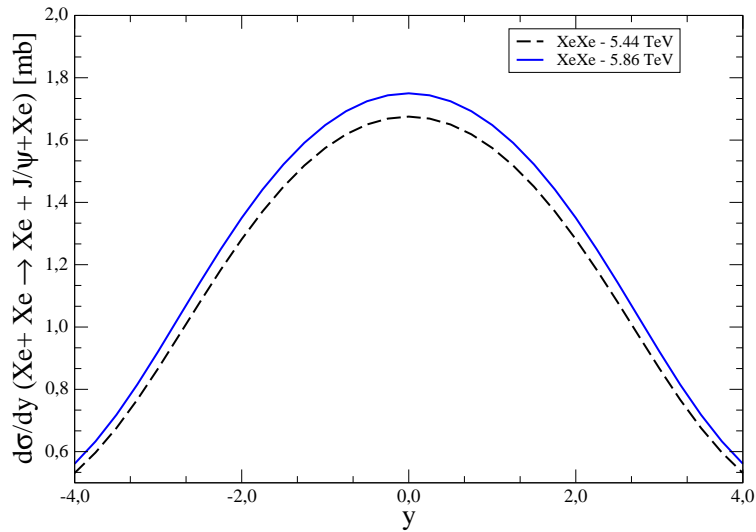}
 \caption{Rapidity distribution of $J/\psi$  production in XeXe collisions at the LHC ($\sqrt{s_{\mathrm{NN}}} = 5.44$ and $5.86$ TeV).}
  \label{fig:XeXe544TeV}
\end{figure}

Motivated by the recent ALICE measurement for coherent $\rho^0$  production in XeXe ultraperipheral collisions (UPC)  at 5.44 TeV \cite{ALICE:2021jnv}, we estimate the rapidity distribution for the $J/\psi$ production at 5.44 and 5.86 TeV. This is presented in Fig. \ref{fig:XeXe544TeV}. In this case the shadowing effects are expected to depend on the $A$ and UPC measurements for different atomic number extend the study of the shadowing. Along these lines was reported performance projections for  measurements that can be carried out with an Oxigen-Oxigen run during the LHC Run 3 \cite{ALICEOO,Brewer:2021kiv}.  In the impulse approximation, $d\sigma_{\gamma A}(t=0)/dt\approx A^2d\sigma_{\gamma p}(t=0)/dt$, and in case of similar energies it is expected that the XeXe cross section be a factor 0.4 smaller than for PbPb collisions.

\begin{figure}[t]
\centering
 \includegraphics[width=.6\textwidth]{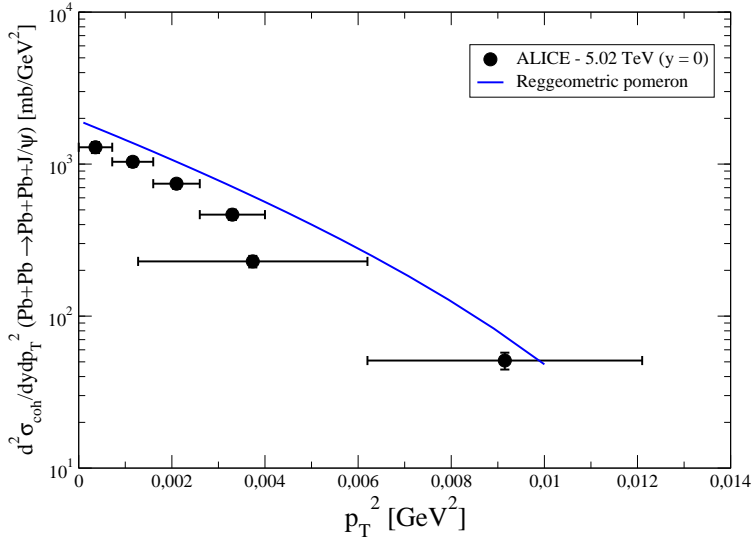}
 \caption{Transverse momentum of $J/\psi$  production in PbPb collisions at the LHC ($\sqrt{s_{\mathrm{NN}}} = 5.02$ TeV). Prediction of  the  Reggeometric Pomeron model is compared to data from ALICE collaboration \cite{ALICE:2021tyx}.}
  \label{fig:dsdpt502TeV}
\end{figure}

Finally, we present our predictions for the transverse momentum distribution for $J/\psi$ production which has been recently measured by ALICE collaboration \cite{ALICE:2021tyx} in PbPb collisions at $\sqrt{s_{\mathrm{NN}}} = 5.02$ TeV. The $p_T$-distribution is obtained theoretically in the following way:
\begin{eqnarray}
\frac{d^2\sigma (A+A\rightarrow A+V+A)}{dydp_T^2}=\omega_+\frac{dN_{\gamma/A}(\omega_+)}{d\omega}\frac{d\sigma_{\gamma A\rightarrow VA}(\omega_+)}{d|t|} +\omega_-\frac{dN_{\gamma/A}(\omega_-)}{d\omega}\frac{d\sigma_{\gamma A\rightarrow VA}(\omega_-)}{d|t|},
\end{eqnarray}
where the $t$-dependence is obtained directly from Eq. (\ref{eq:sigtotgammaA}) by removing the integration in $|t|$. In Fig. \ref{fig:dsdpt502TeV} the result is presented and compared to ALICE data at $y=0$. As it can be seen, the Reggeometric Pomeron model overestimates the data in consistency with the rapidity distribution at mid-rapidity in Fig. \ref{fig:VMAA502TeV}. Similar overestimation is also observed for the prediction from STARLIGHT Monte Carlo, where the  shape of the $|t|$-distribution is wider than that of experimental measurement. The normalization and shape  of data seem to be better described with the leading-twist approximation (LTA) of nuclear shadowing considering a low shadowing prediction (see Ref. \cite{Guzey:2020ntc}). 


\section{Conclusions}

In this paper predictions for exclusive $J/\Psi$ and $\psi (2S)$ meson production in ultra-peripheral collisions at the LHC are presented and compared with the recent experimental data collected by different collaborations. We update our previous studies in Refs. \cite{Fiore:2014oha,Fiore:2014lxa,Fiore:2015yya}  by analysing the measurements in proton-proton collisions at 13 TeV. The cross section $\sigma (\gamma+p\rightarrow J/\psi+p)$ is described in terms of the single-component Reggeometric Pomeron model. It describes nicely the recent indirect measurements (see Fig. \ref{fig:energydependence})  and the HERA  experimental data. A better description can be obtained by further tuning the power. The LHCb data at 13 TeV are described reasonably given the band of uncertainty for the model. The description of $\psi (2S)$ data is good. By using the Glauber model and VDM approach, the predictions for PbPb collisions at 2.76 and 5.02 TeV are presented. Concerning the rapidity distributions, the model tends to describe better the forward rapidity region and overestimates the data at central rapidity. This feature is similar to the approaches that not consider leading-twist  nuclear shadowing. Prediction for XeXe UPC collision has been provided. We provide prediction for the transverse momentum distribution and compared it to the first measurement of $|t|$-distribution performed by ALICE  collaboration \cite{ALICE:2021tyx}. 

The present study can be extended by:
\begin{itemize}
\item the inclusion of the $t$ dependence of the differential cross section with deviations from the usual linear parameterization (non-linear Regge trajectories) \cite{Fazio:2011ex};
\item the inclusion of other models beyond the single-component Reggeometric Pomeron like the two-pomeron models and the soft dipole pomeron \cite{FFJS2};
\item studies on incoherent vector meson production, $A+A \rightarrow A+V+A^*$, in ultraperipheral heavy ion collisions \cite{Klein:2016yzr,Lappi:2013am,Toll:2013gda,Strikman:2005ze,Cepila:2017nef,Traini:2018hxd};
\item studies on production at low transverse momentum and its dependence on centrality in PbPb collisions \cite{LHCb:2021hoq}. 
\end{itemize}

 \section*{Acknowledgements}
L.J. was supported by the grant 1230 \textit{Fundamental properties of
matter in relativistic nuclear collisions and early Universe} of the
National Academy of Sciences of Ukraine, 0120U100935.
MVTM was supported by funding agencies CAPES and CNPq (Brazil).


\begin{thebibliography}{99}

\bibitem{ALICE:2019tqa}
S.~Acharya \textit{et al.} [ALICE],
Phys. Lett. B \textbf{798}, 134926 (2019)
doi:10.1016/j.physletb.2019.134926
[arXiv:1904.06272 [nucl-ex]].

\bibitem{ALICE:2021gpt}
S.~Acharya \textit{et al.} [ALICE],
Eur. Phys. J. C \textbf{81}, no.8, 712 (2021)
doi:10.1140/epjc/s10052-021-09437-6
[arXiv:2101.04577 [nucl-ex]].

\bibitem{LHCb:2021bfl}
R.~Aaij \textit{et al.} [LHCb],
[arXiv:2107.03223 [hep-ex]].


\bibitem{ALICE:2021tyx}
S.~Acharya \textit{et al.} [ALICE],
Phys. Lett. B \textbf{817}, 136280 (2021)
doi:10.1016/j.physletb.2021.136280
[arXiv:2101.04623 [nucl-ex]].

\bibitem{LHCb:2021hoq}
R.~Aaij \textit{et al.} [LHCb],
[arXiv:2108.02681 [hep-ex]].

\bibitem{Review} G. Baur {\it et al.}, Phys. Rept. {\bf 364} (2002) 359; K. Hencken {\it et al.}, Phys. Rept. {\bf 458} (2008) 1.

\bibitem{Contreras:2015dqa}
J.~G.~Contreras and J.~D.~Tapia Takaki,
Int. J. Mod. Phys. A \textbf{30}, 1542012 (2015)
doi:10.1142/S0217751X15420129

\bibitem{Schafer2020} W. Sch\"{a}fer, Eur. Phys. J. A \textbf{56}, 231 (2020). 

\bibitem{Klein:2020fmr}
S.~Klein and P.~Steinberg,
Ann. Rev. Nucl. Part. Sci. \textbf{70}, 323-354 (2020)
doi:10.1146/annurev-nucl-030320-033923
[arXiv:2005.01872 [nucl-ex]].

\bibitem{Klein:1999qj}
S.~Klein and J.~Nystrand,
Phys. Rev. C \textbf{60}, 014903 (1999)
doi:10.1103/PhysRevC.60.014903
[arXiv:hep-ph/9902259 [hep-ph]].

\bibitem{Guzey:2016piu}
V.~Guzey, E.~Kryshen and M.~Zhalov,
Phys. Rev. C \textbf{93}, no.5, 055206 (2016)
doi:10.1103/PhysRevC.93.055206
[arXiv:1602.01456 [nucl-th]].

\bibitem{Klusek-Gawenda:2015hja}
M.~K\l{}usek-Gawenda and A.~Szczurek,
Phys. Rev. C \textbf{93}, no.4, 044912 (2016)
doi:10.1103/PhysRevC.93.044912
[arXiv:1509.03173 [nucl-th]].

\bibitem{Bautista:2016xnp}
I.~Bautista, A.~Fernandez Tellez and M.~Hentschinski,
Phys. Rev. D \textbf{94}, no.5, 054002 (2016)
doi:10.1103/PhysRevD.94.054002
[arXiv:1607.05203 [hep-ph]].

\bibitem{Hentschinski:2020yfm}
M.~Hentschinski and E.~Padr\'on Molina,
Phys. Rev. D \textbf{103}, no.7, 074008 (2021)
doi:10.1103/PhysRevD.103.074008
[arXiv:2011.02640 [hep-ph]].

\bibitem{Jones:2013pga}
S.~P.~Jones, A.~D.~Martin, M.~G.~Ryskin and T.~Teubner,
JHEP \textbf{11}, 085 (2013)
doi:10.1007/JHEP11(2013)085
[arXiv:1307.7099 [hep-ph]].

\bibitem{Jones:2013eda}
S.~P.~Jones, A.~D.~Martin, M.~G.~Ryskin and T.~Teubner,
J. Phys. G \textbf{41}, 055009 (2014)
doi:10.1088/0954-3899/41/5/055009
[arXiv:1312.6795 [hep-ph]].

\bibitem{Goncalves:2017wgg}
V.~P.~Gon\c{c}alves, M.~V.~T.~Machado, B.~D.~Moreira, F.~S.~Navarra and G.~S.~dos Santos,
Phys. Rev. D \textbf{96}, no.9, 094027 (2017)
doi:10.1103/PhysRevD.96.094027
[arXiv:1710.10070 [hep-ph]].

\bibitem{SampaiodosSantos:2014puz}
G.~Sampaio dos Santos and M.~V.~T.~Machado,
J. Phys. G \textbf{42}, no.10, 105001 (2015)
doi:10.1088/0954-3899/42/10/105001
[arXiv:1411.7918 [hep-ph]].

\bibitem{Lappi:2010dd}
T.~Lappi and H.~Mantysaari,
Phys. Rev. C \textbf{83}, 065202 (2011)
doi:10.1103/PhysRevC.83.065202
[arXiv:1011.1988 [hep-ph]].

\bibitem{Lappi:2013am}
T.~Lappi and H.~Mantysaari,
Phys. Rev. C \textbf{87}, no.3, 032201 (2013)
doi:10.1103/PhysRevC.87.032201
[arXiv:1301.4095 [hep-ph]].

\bibitem{Cepila:2016uku}
J.~Cepila, J.~G.~Contreras and J.~D.~Tapia Takaki,
Phys. Lett. B \textbf{766}, 186-191 (2017)
doi:10.1016/j.physletb.2016.12.063
[arXiv:1608.07559 [hep-ph]].

\bibitem{Cepila:2017nef}
J.~Cepila, J.~G.~Contreras and M.~Krelina,
Phys. Rev. C \textbf{97}, no.2, 024901 (2018)
doi:10.1103/PhysRevC.97.024901
[arXiv:1711.01855 [hep-ph]].

\bibitem{Luszczak:2019vdc}
A.~\L{}uszczak and W.~Sch\"afer,
Phys. Rev. C \textbf{99}, no.4, 044905 (2019)
doi:10.1103/PhysRevC.99.044905
[arXiv:1901.07989 [hep-ph]].

\bibitem{GayDucati:2018who}
M.~B.~Gay Ducati and S.~Martins,
Phys. Rev. D \textbf{97}, no.11, 116013 (2018)
doi:10.1103/PhysRevD.97.116013
[arXiv:1804.09836 [hep-ph]].

\bibitem{Jenkovszky:2018itd}
L.~Jenkovszky, R.~Schicker and I.~Szanyi,
Int. J. Mod. Phys. E \textbf{27}, no.08, 1830005 (2018)
doi:10.1142/S0218301318300059
[arXiv:1902.05614 [hep-ph]].

\bibitem{Fiore:2014oha}
R.~Fiore, L.~Jenkovszky, V.~Libov and M.~Machado,
Teor. Mat. Fiz. \textbf{182}, no.1, 171-181 (2014)
doi:10.1007/s11232-015-0252-8
[arXiv:1408.0530 [hep-ph]].

\bibitem{Fiore:2014lxa}
R.~Fiore, L.~Jenkovszky, V.~Libov, M.~V.~T.~Machado and A.~Salii,
[arXiv:1506.01990 [hep-ph]]. Contribution to: Diffraction 2014.

\bibitem{Fiore:2015yya}
R.~Fiore, L.~Jenkovszky, V.~Libov, M.~V.~T.~Machado and A.~Salii,
AIP Conf. Proc. \textbf{1654}, no.1, 090002 (2015)
doi:10.1063/1.4916009

\bibitem{Capua} M. Capua {\it et al.}, Phys. Lett. {\bf B645} (1997) 161, hep-ph/0605319.

\bibitem{FFJS1} S.~Fazio, R.~Fiore, A.~Lavorini, L.~Jenkovszky and A.~Salii,
Acta Phys. Polon. B \textbf{44}, 1333-1353 (2013),
[arXiv:1304.1891 [hep-ph]].

\bibitem{FFJS2} S. Fazio, R. Fiore, L. Jenkovszky, and A. Salii, Phys. Rev. D \textbf{90}, no.1, 016007 (2014),
[arXiv:1312.5683 [hep-ph]].


\bibitem{ALICE:2014eof}
B.~B.~Abelev \textit{et al.} [ALICE],
Phys. Rev. Lett. \textbf{113}, no.23, 232504 (2014)
doi:10.1103/PhysRevLett.113.232504
[arXiv:1406.7819 [nucl-ex]].


\bibitem{ALICE:2018oyo}
S.~Acharya \textit{et al.} [ALICE],
Eur. Phys. J. C \textbf{79}, no.5, 402 (2019)
doi:10.1140/epjc/s10052-019-6816-2
[arXiv:1809.03235 [nucl-ex]].

\bibitem{LHCb:2018rcm}
R.~Aaij \textit{et al.} [LHCb],
JHEP \textbf{10}, 167 (2018)
doi:10.1007/JHEP10(2018)167
[arXiv:1806.04079 [hep-ex]].


\bibitem{Jones:2016icr}
S.~P.~Jones, A.~D.~Martin, M.~G.~Ryskin and T.~Teubner,
J. Phys. G \textbf{44}, no.3, 03LT01 (2017)
doi:10.1088/1361-6471/aa56ea
[arXiv:1611.03711 [hep-ph]].

\bibitem{H1:2013okq}
C.~Alexa \textit{et al.} [H1],
Eur. Phys. J. C \textbf{73}, no.6, 2466 (2013)
doi:10.1140/epjc/s10052-013-2466-y
[arXiv:1304.5162 [hep-ex]].


\bibitem{Klein:2016yzr}
S.~R.~Klein, J.~Nystrand, J.~Seger, Y.~Gorbunov and J.~Butterworth,
Comput. Phys. Commun. \textbf{212}, 258-268 (2017)
doi:10.1016/j.cpc.2016.10.016
[arXiv:1607.03838 [hep-ph]].
 

\bibitem{ZEUS:2016awm}
H.~Abramowicz \textit{et al.} [ZEUS],
Nucl. Phys. B \textbf{909}, 934-953 (2016)
doi:10.1016/j.nuclphysb.2016.06.010
[arXiv:1606.08652 [hep-ex]].


\bibitem{Bauer:1977iq}
T.~H.~Bauer, R.~D.~Spital, D.~R.~Yennie and F.~M.~Pipkin,
Rev. Mod. Phys. \textbf{50}, 261 (1978)
[erratum: Rev. Mod. Phys. \textbf{51}, 407 (1979)]
doi:10.1103/RevModPhys.50.261

\bibitem{Frankfurt:2003wv}
L.~Frankfurt, M.~Strikman and M.~Zhalov,
Acta Phys. Polon. B \textbf{34}, 3215-3254 (2003)
[arXiv:hep-ph/0304301 [hep-ph]].

\bibitem{Davies:1976zzb}
K.~T.~R.~Davies and J.~R.~Nix,
Phys. Rev. C \textbf{14}, 1977-1994 (1976)
doi:10.1103/PhysRevC.14.1977

\bibitem{ALICE:2013wjo}
E.~Abbas \textit{et al.} [ALICE],
Eur. Phys. J. C \textbf{73}, no.11, 2617 (2013)
doi:10.1140/epjc/s10052-013-2617-1
[arXiv:1305.1467 [nucl-ex]].

\bibitem{ALICE:2012yye}
B.~Abelev \textit{et al.} [ALICE],
Phys. Lett. B \textbf{718}, 1273-1283 (2013)
doi:10.1016/j.physletb.2012.11.059
[arXiv:1209.3715 [nucl-ex]].

\bibitem{ALICE:2015nmy}
J.~Adam \textit{et al.} [ALICE],
Phys. Lett. B \textbf{751}, 358-370 (2015)
doi:10.1016/j.physletb.2015.10.040
[arXiv:1508.05076 [nucl-ex]].

\bibitem{CMS:2016itn}
V.~Khachatryan \textit{et al.} [CMS],
Phys. Lett. B \textbf{772}, 489-511 (2017)
doi:10.1016/j.physletb.2017.07.001
[arXiv:1605.06966 [nucl-ex]].

\bibitem{Guzey:2013xba}
V.~Guzey, E.~Kryshen, M.~Strikman and M.~Zhalov,
Phys. Lett. B \textbf{726}, 290-295 (2013)
doi:10.1016/j.physletb.2013.08.043
[arXiv:1305.1724 [hep-ph]].

\bibitem{Guzey:2020ntc}
V.~Guzey, E.~Kryshen, M.~Strikman and M.~Zhalov,
Phys. Lett. B \textbf{816}, 136202 (2021)
doi:10.1016/j.physletb.2021.136202
[arXiv:2008.10891 [hep-ph]].



\bibitem{Fazio:2011ex}
S.~Fazio, R.~Fiore, L.~Jenkovszky and A.~Lavorini,
Phys. Rev. D \textbf{85}, 054009 (2012)
doi:10.1103/PhysRevD.85.054009
[arXiv:1109.6374 [hep-ph]].



\bibitem{Toll:2013gda}
T.~Toll and T.~Ullrich,
Comput. Phys. Commun. \textbf{185}, 1835-1853 (2014)
doi:10.1016/j.cpc.2014.03.010
[arXiv:1307.8059 [hep-ph]].

\bibitem{Strikman:2005ze}
M.~Strikman, M.~Tverskoy and M.~Zhalov,
Phys. Lett. B \textbf{626}, 72-79 (2005)
doi:10.1016/j.physletb.2005.08.083
[arXiv:hep-ph/0505023 [hep-ph]].




\bibitem{Traini:2018hxd}
M.~C.~Traini and J.~P.~Blaizot,
Eur. Phys. J. C \textbf{79}, no.4, 327 (2019)
doi:10.1140/epjc/s10052-019-6826-0
[arXiv:1804.06110 [hep-ph]].

\bibitem{ALICE:2021jnv}
S.~Acharya \textit{et al.} [ALICE],
Phys. Lett. B \textbf{820}, 136481 (2021)
doi:10.1016/j.physletb.2021.136481
[arXiv:2101.02581 [nucl-ex]].

\bibitem{ALICEOO} ALICE Collaboration, \textit{ALICE physics projections for a short oxygen-beam run at the LHC}, ALICE-PUBLIC-2021-004.

\bibitem{Brewer:2021kiv}
J.~Brewer, A.~Mazeliauskas and W.~van der Schee,
[arXiv:2103.01939 [hep-ph]].


\end{thebibliography}
\end{document}